\documentclass[9pt,twocolumn,twoside]{opticajnl}
\journal{opticajournal} % use for journal or Optica Open submissions

\setboolean{shortarticle}{false}
\usepackage{lineno}
\usepackage{layouts}
\usepackage[per-mode=repeated-symbol]{siunitx}
\title{XUV yield optimization of two-color high-order harmonic generation in gases}

\author[1,*]{Ann-Kathrin Raab}
\author[1]{Melvin Redon}
\author[2]{Sylvianne Roscam Abbing}
\author[1]{Yuman Fang}
\author[1]{Chen Guo}
\author[2]{Peter Smorenburg}
\author[1]{Johan Mauritsson}
\author[1]{Anne-Lise Viotti}
\author[1]{Anne L'Huillier}
\author[1]{Cord L. Arnold}

\affil[1]{Department of Physics, Lund University, P.O. Box 118, 22100, Lund, Sweden}
\affil[2]{ASML Research, ASML Netherlands B.V., 5504 DR, Veldhoven, Netherlands}

\affil[*]{ann-kathrin.raab@fysik.lth.se}

\begin{abstract}
We perform an experimental two-color high-order harmonic generation study in argon with the fundamental of an ytterbium ultrashort pulse laser and its second harmonic. The intensity of the second harmonic and its phase relative to the fundamental are varied, in a large range compared to earlier works, while keeping the total intensity constant. We extract the optimum values for the relative phase and ratio of the two colors which lead to a maximum yield enhancement for each harmonic order in the extreme ultraviolet spectrum. Within the semi-classical three-step model, the yield maximum can be associated with a flat electron return time vs.~return energy distribution.
An analysis of different distributions allows to predict the required relative two-color phase and ratio for a given harmonic order, total laser intensity, fundamental wavelength, and ionization potential.
\end{abstract}

\setboolean{displaycopyright}{false} % Do not include copyright or licensing information in submission.

\begin{document}

\maketitle

\section{Introduction}
High-order harmonic generation (HHG) with more than one color is a very versatile tool. 
By overlapping the fundamental laser driver with, usually,  its second or third harmonic, 
%the acceleration and recombination behavior of electrons resulting in 
the properties of the high-order harmonics can be optimized depending on the individual driving color fields' ratio and phase relationship. 
Experimentally, spatial \cite{roscamabbing_divergence_2021}, spectral \cite{mansten_spectral_2008, wei_selective_2013, wei_efficient_2014, mitra_suppression_2020}, and temporal \cite{mauritsson_attosecond_2006} shaping of the resulting extreme ultraviolet (XUV) radiation have already been demonstrated by using multicolor driving fields. It has also served as a tool to better understand the fundamentals of HHG \cite{dudovich_measuring_2006, shafir_atomic_2009,fiess_attosecond_2011}.
One of the most prevalent properties of two-color HHG is the increase in yield compared to single-color HHG, with up to several orders of magnitude enhancement \cite{watanabe_twocolor_1994, kim_highly_2005, liu_significant_2006, ganeev_enhancement_2009,brugnera_enhancement_2010, roscamabbing_divergence_2021}. 

The science behind XUV yield optimization becomes increasingly important with HHG entering the industrial realms as e.g.~metrology tool \cite{porterSoftXrayNovel2023}. Two different aspects are to be considered, the single-atom response, which can be intuitively described by the semi-classical three-step model \cite{corkum_plasma_1993, schafer_threshold_1993}, as well as propagation effects in the nonlinear medium \cite{constant_optimizing_1999, weissenbilderHow2022}. In the three-step model, an electron first tunnels out through the potential distorted by the laser field (step one), travels in the continuum driven by the strong electric field (step two), before it eventually recombines with the parent atom (step three). The energy of the XUV photon that is emitted then corresponds to its return kinetic energy $E_R$ plus the ionization potential $I_p$ of the atom. When using a multi-color driver, both the ionization rate and the trajectory taken by an electron are affected by the change in the electric field shape.
%the ionization rate is affected due to the wavelength and field dependence \cite{ammosovTunnelIonizationComplex1986} as well as the trajectory that an electron takes due to the change in the electric field shape.  
Theoretical studies have provided waveform parameters for multi-color synthesis to enhance flux and/or cutoff \cite{chipperfield_ideal_2009, jinRouteOptimalGeneration2014, jinWaveformsOptimalSubkeV2014}, or to optimize  phase-matching \cite{schiesslEnhancementHighorderHarmonic2006, chenEffect2023, stremoukhovRole2023}, all based on relatively complex simulations. The objective of the study presented in this work is to provide a simple rule giving the optimum relative phase and ratio, to enhance the XUV yield of a specific harmonic order, using a two-color field consisting of the fundamental and its second harmonic for a given total intensity and gas species. We hereby use, verify, and extend the findings of Raz et al., connecting local extrema in the return energy vs.~ionization time distribution to yield enhancements \cite{raz_spectral_2012}, based on predictions from catastrophe theory \cite{berryIVCatastropheOptics1980}.

\begin{figure*}[t]
    \centering
    \includegraphics[]{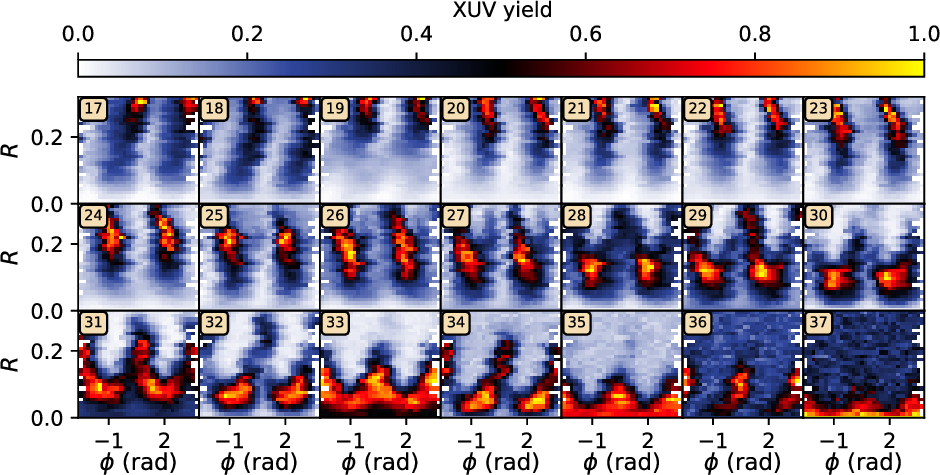}
    \caption{XUV yield for individual harmonic orders, depending on $R$ and $\phi$ for $I_{\text{tot}}=\SI{0.9e14}{\watt\per\centi\meter\squared}$. For each harmonic, with the order indicated by the numbered box, the yield is normalized to its maximum.}
    \label{fig:results}
\end{figure*}
\section{Experiment}

\subsection{Setup}
The experimental setup consists of an interferometric two-color synthesizer, where part of the fundamental ($\omega$, $\SI{1030}{\nano\meter}$) is mixed with its second harmonic ($2\omega$, $\SI{515}{\nano\meter}$). The synthesizer is driven by an ytterbium laser system (Pharos, Light Conversion) at $\SI{10}{\kilo\hertz}$ repetition rate with a maximum pulse energy of $\SI{700}{\micro\joule}$ and a pulse duration of $\SI{180}{\femto\second}$. In each interferometer arm, a spatial light modulator (SLM) is used to shape the spatial phase to correct for aberrations, perform fine alignment, and adjust the relative phase $\phi$ in a $2\pi$ range. In addition, a thin film polarizer and a motorized wave plate vary the relative contribution of the $\omega$ and $2\omega$ fields, determined by

\begin{equation}\label{eq:theory:R}
    R = \frac{I_{2\omega}}{I_{2\omega}+I_{\omega}} = \frac{I_{2\omega}}{I_{\text{tot}}} \, ,
\end{equation}
where $I_{\omega}$, $I_{2\omega}$ and $I_{\text{tot}}$ denote the intensity of the fundamental, the second harmonic and the total intensity. 
Both colors are recombined and focused with a singlet lens ($f=\SI{250}{\milli\meter}$) into an argon gas target, consisting of a gas jet from a nozzle with a $\SI{42}{\micro\meter}$ diameter and a backing pressure of $\SI{2}{\bar}$. The focused beams have a radius of $w_{\omega}=\SI{27}{\micro\meter}$ and $w_{2\omega}=\SI{14}{\micro\meter}$, respectively.
The total electric field in the gas target is defined by
\begin{equation}\label{eq:theory:twocolorE}
\begin{aligned}
     E(t) = E_{\omega} +E_{2\omega} &=\frac{1}{\sqrt{c\epsilon_0}}\left[\sqrt{(1-R)I_{\text{tot}}}\cos(\omega t) \right. \\ 
     &+ \left. \sqrt{RI_{\text{tot}}}\cos(2\omega t+\phi)\right] \, ,
\end{aligned}
 %   E(t)=\frac{1}{\sqrt{c\epsilon_0}}\left[\sqrt{(1-R)I_{\text{tot}}}\cos(\omega t) + \sqrt{RI_{\text{tot}}}\cos(2\omega t+\phi)\right] \, , 
\end{equation}
with the speed of light $c$ and the vacuum permittivity $\epsilon_0$. Both colors are linearly polarized parallel to each other.
The generated XUV radiation is sent on a curved grating and detected by a multi-channel plate with a phosphor screen behind, which is imaged by a camera. The  experimental setup including the two-color field  diagnostics is described in detail in a previous publication \cite{raabHighlyVersatileTwocolor2024}.

\subsection{Results}
In Fig.~\ref{fig:results} the XUV yield for the harmonic orders 17 to 37 is shown as a function of $R$ and $\phi$, for a constant total intensity of $I_{\text{tot}}=\SI{0.9e14}{\watt\per\centi\meter\squared}$ in identical experimental conditions. Each data point consists of the integrated counts for one harmonic order. 
Due to minor phase instabilities, $\phi$ is measured every $\SI{200}{\milli\second}$ which is used to sort and interpolate the data to a common grid with respect to $R$ and $\phi$, as shown in Fig.~\ref{fig:results}.
%During the experiment,  $\phi$ is measured every $\SI{200}{\milli\second}$ to reinterpolate the count rates to the phase axis shown in Fig.~\ref{fig:results}. 
No extrapolation towards the limits of the $\phi$-axis is performed, causing the white regions in all subfigures.  
For illustration purposes, each subfigure corresponding to one harmonic order is normalized to its maximum yield. With increasing harmonic order, the $R$-value leading to the maximum yield, $R_{\text{opt}}$, decreases.

\subsection{Analysis}
\begin{figure}[t]
    \centering
    \includegraphics{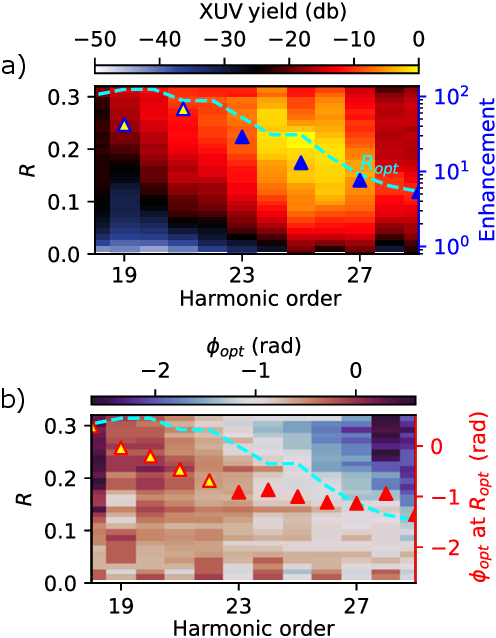}
    \caption{a): $A+C$ from \eqref{eq:2color:fit} for $R$ and harmonic order (color) within the plateau ($H=18-29$). The optimum ratio $R_{\text{opt}}$ (cyan dashed line) marks the maximum yield for each harmonic order. For odd orders, the enhancement of the yield compared to $R=0$ is given on the right axis (blue triangles). b): $\phi_{\text{opt}}$ for $R$ and harmonic order (color, cyclic scale). The $\phi_{\text{opt}}$ value at $R_{\text{opt}}$ below the cyan dashed line is extracted and shown on the right axis (red triangles). The triangles filled with yellow $H<23$ denote that a global yield maximum was not reached in the experiment.}
    \label{fig:fit}
\end{figure}

To determine  $R_{\text{opt}}$ as well as the optimum phase  $\phi_{\text{opt}}$ more accurately, a fit was conducted for every line of experimental data with respect to $\phi$ according to
\begin{equation}\label{eq:2color:fit}
    f_{R,H}(\phi) = A\cos(2(\phi-\phi_{\text{opt}})) + C \, .
\end{equation}
$A$ represents the modulation amplitude with $\phi$ of an harmonic order $H$, while $C$ represents the constant signal baseline. The measure $A+C$ then represents the maximum yield at $\phi_{\text{opt}}$. In Fig.~\ref{fig:fit} a), $A+C$ is shown as a color scale for all harmonic orders within the HHG plateau ($H=18-29$), as a function of $R$.  The cyan dashed line represents $R_{\text{opt}}$. The right axis represents the yield enhancement, which is calculated by dividing the yield at $R_{\text{opt}}$ by the yield at $R=0$, for odd orders, shown as blue triangles.
For $H<23$, a global yield maximum at $R_{\text{opt}}$ could not be reached due to the limitation for $R$ in the experiment; see Fig.~\ref{fig:results}. The triangles filled with yellow thus denote the enhancement that was reached at $R=0.32$.

Figure \ref{fig:fit} b) shows $\phi_{\text{opt}}$ for all harmonic orders as a function of $R$ on a cyclic color scale. Overlaying the dashed cyan line ($R_{\text{opt}}$), it becomes evident that the phase values corresponding to this line seem to be constant (white color). These values of $\phi_{\text{opt}}$ at $R_{\text{opt}}$ are extracted and shown on the right axis as red triangles. As soon as a global yield optimum in Fig.~\ref{fig:results} is reached from $H=23$ on, $\phi_{\text{opt}}$ at $R_{\text{opt}}$ indeed is concentrated around $\phi_{\text{opt}} = \SI{-1}{\radian}$. This applies only for $H>22$, where a global maximum of the yield could be reached. For $H<23$ (yellow-filled triangles) the optimum phase at the yield maximum at $R=0.32$ is shown instead.

\section{Discussion}
\subsection{Three-step model at $\phi = \SI{-1}{\radian}$}
\begin{figure}
    \centering
    \includegraphics{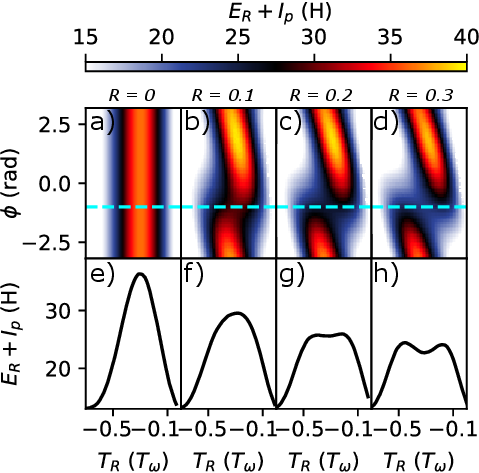}
    \caption{a)-d): $E_R+I_p$ in the unit of harmonic order $H$ for electrons ionized during one electric field half cycle, as a function of $\phi$ and return time $T_R$ in units of the electric field cycle of the fundamental $T_{\omega}$, for four examples of $R$. The cyan line marks the experimentally discovered optimum phase ${\phi_{\text{opt}} = \SI{-1}{\radian}}$. e)-h): The energy distribution at the position of the cyan lines as line-outs.}
    \label{fig:trajectories1}
\end{figure}

To understand what is happening at $\phi_{\text{opt}} = \SI{-1}{\radian}$, the three-step model is evaluated, calculating the electron trajectories for an electric field according to \eqref{eq:theory:twocolorE}, for varying ionization times $T_I$ and extracting the kinetic energy and the return time $T_R$ of the electrons that recombine with the parent atom, matching the conditions of the experiment with $I_{\text{tot}}=\SI{0.9e14}{\watt\per\centi\meter\squared}$. 
In Fig.~\ref{fig:trajectories1} a)-d), the corresponding XUV photon energy with ${I_p=\SI{15.7}{\electronvolt}}$ is shown for different values of $R$ as a function of $\phi$ and $T_R$ in units of the single-color laser cycle $T_{\omega}$, for one electric field half-cycle. 
%As $\phi$ is varied in a $2\phi$ range, the effect on the second half-cycle return energy distribution corresponds to the same plot, but is shifted by $\pi$.
$E_R+I_p$ is given in units of the corresponding harmonic order $H$. For $R=0$ (a), which corresponds to a standard single-color field, there is no variation with $\phi$ and the electrons in the two half-cycles behave identically with respect to their return energy. When $R$ increases, from a) to d) in Fig.~\ref{fig:trajectories1}, the return energies of the electrons in the first half cycle are suppressed, while they are enhanced for the second one. The location of the largest suppression corresponds to $\phi=\SI{-1}{\radian}$, which is indicated by the cyan dashed line. The distributions of $E_R+I_p$ corresponding to this line are shown in Fig.~\ref{fig:trajectories1} e)-h) as line-outs.
The essential observation is that this suppression of return energy is what causes an increase of harmonic yield.

The harmonic yield can be estimated by assuming that it is proportional to $(dE_R/dT_I)^{-1}$ \cite{corkum_plasma_1993}. This expression diverges for a maximum, such as the  cutoff at $3.17U_p$ for a one-color field with 
\begin{equation}
    U_p=\frac{e^2}{4 m_{\text{e}}\sqrt{c\epsilon_0}} \cdot \frac{I}{\omega^2} \, .
\end{equation}
Here, $e$ is the electron charge and $m_{\text{e}}$ the electron mass.  This situation corresponds to the $R=0$ line-out in Fig.~\ref{fig:trajectories1} which leads to a rapid decrease of the strength of the harmonic orders in the cutoff regime. Such a singularity can also be treated with catastrophe theory \cite{berryIVCatastropheOptics1980} in the form of caustics. 
Caustics in the context of optics describe high concentrations of intensity in a diffraction pattern \cite{kravtsovCausticsCatastrophesWave1983}. In the context of two-color HHG, the enhancement of yield is proportional to the order of the return energy vs.~return time singularity, which is related to how many orders of derivatives of the curves shown in  Fig.~\ref{fig:trajectories1} e)-h) vanish \cite{raz_spectral_2012}.
Thus, addition of a second harmonic field can suppress the return energies in a certain range of the optical cycle, which flattens the time-energy curve. This makes electrons coalesce to the same energy of the emitted harmonic, which is thus enhanced.

\begin{figure}[t]
    \centering
    \includegraphics{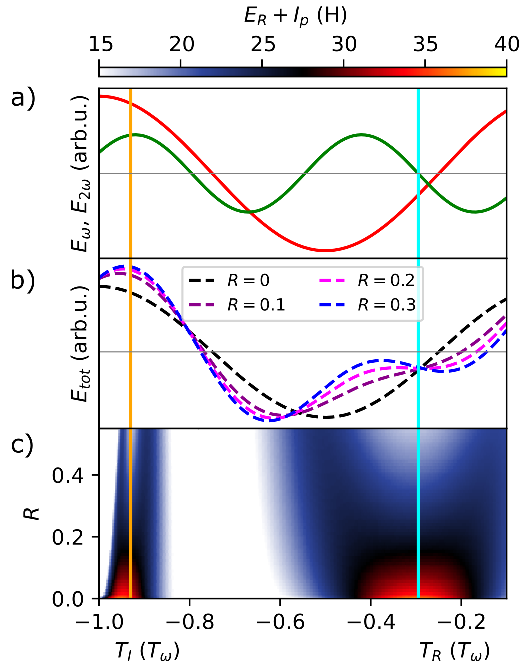}
    \caption{a): Electric field for $\omega$ (red) and $2\omega$ (green) for ${\phi = \SI{-1}{\radian}}$ and $R=0.2$. b): Total electric field according to \eqref{eq:theory:twocolorE} for different values of $R$ and $\phi = \SI{-1}{\radian}$. c): $E_R$ + $I_p$ at the corresponding ionization time $T_I$ and return time $T_R$ at $\phi = \SI{-1}{\radian}$, sharing the color scale of Fig.~\ref{fig:trajectories1}. The approximately constant flattening location with respect to ionization (orange) and return (cyan) is highlighted with solid vertical lines.}
    \label{fig:trajectories2}
\end{figure}

To calculate the harmonic order to which the location of the time-energy curve flattening corresponds, the contributions of $\omega$ and $2\omega$ are considered separately.
In Fig.~\ref{fig:trajectories2} a), the fields are plotted for $\phi=\SI{-1}{\radian}$ at an exemplary ratio of $R=0.2$  for $\omega$ (red) and $2\omega$ (green). 
In Fig.~\ref{fig:trajectories2} b), the resulting total electric field according to \eqref{eq:theory:twocolorE} for $\phi=\SI{-1}{\radian}$ and different $R$ values is shown. The electric fields hereby correspond to the electron return energy distributions highlighted in the e)-h) of Fig.~\ref{fig:trajectories1}.
Figure \ref{fig:trajectories2} c) shows the calculated XUV photon energies ($E_R+I_p$) in the unit of the corresponding harmonic order, sharing the same color scale as Fig.~\ref{fig:trajectories1}, as a function of the ionization $T_I$ and return time $T_R$ at $\phi = \SI{-1}{\radian}$. 

The electrons that contribute to the flattening of the time-energy curve are ionized and return approximately at the same times, independently of $R$. This is indicated by the vertical lines for ionization (orange, $T_I=-0.93\, T_{\omega}$) and recombination (cyan,  $T_R=-0.295\, T_{\omega}$). 
The total field plotted in Fig.~\ref{fig:trajectories2} b) is approximately highest at the time of ionization. At the time of recombination the $2\omega$ contribution is equal to zero, as highlighted in Fig.~\ref{fig:trajectories2} a). Interestingly, the time of return coincides with an isosbestic point of the fields, independent of $R$. This means that the electric field $E(t=T_R,\phi=\SI{-1}{\radian},R)$ upon return has the same amplitude. For $R=0.2$, which is associated with the highest enhancement of the harmonic order corresponding to the plateau in the time-energy curve (see Fig.~\ref{fig:trajectories1} g)) the first derivative of $E_{\text{tot}}$ also vanishes (pink dashed line). 

%These features remain true for $R<1$.

The absolute value of the velocity $v$ an electron acquires upon recombination is calculated by
\begin{equation}\label{eq:2color:Eintegral}
    v =\frac{p}{m_{\text{e}}} = -\frac{e}{m_{\text{e}}} \int_{T_I}^{T_R} E_{\omega}(t) dt - \frac{e}{m_{\text{e}}} \int_{T_I}^{T_R} E_{2\omega}(t)\, ,
\end{equation}
with $p$ being the momentum and $E_{\omega}$, $E_{2\omega}$ given by \eqref{eq:theory:twocolorE}.
For the fundamental $\omega$, the orange and cyan lines in Fig.~\ref{fig:trajectories2} approximately coincide with the times following the classical cutoff energy, with
\begin{equation}\label{eq:omega_contribution}
 \Delta v_\omega =-\sqrt{\frac{2}{m_{\text{e}}}\cdot3.17 U_p \cdot (1-R)} \, ,
\end{equation}
applying energy conservation and $E_R=m_{\text{e}}v^2/2$.
For ${R=0}$, \eqref{eq:omega_contribution} corresponds to the well-known cutoff law \cite{corkum_plasma_1993}.
For the $2\omega$ component, the vertical lines coincide with the field peak (ionization) and the zero-crossing (recombination), greatly simplifying the evaluation of \eqref{eq:2color:Eintegral} and leading to
\begin{equation}
    \Delta v_{2\omega} = \frac{e\sqrt{RI_{\text{tot}}}}{m_{\text{e}} 2 \omega} = \sqrt{\frac{RU_p}{m_{\text{e}}}} \, ,
\end{equation}
using the definition of the ponderomotive potential $U_p$ for ${R=0}$.

The XUV photon energy corresponding to electrons at the flat region of $E_R$ vs.~$T_R$ is expressed as
\begin{equation}\label{eq:2color:master_theory}
\begin{aligned}
    E_H &= I_p + \frac{1}{2}m_{\text{e}}(\Delta v_\omega + \Delta v_{2\omega})^2 \\ &= I_p + \frac{1}{2}U_p\left( \sqrt{2\cdot 3.17\cdot (1-R)} - \sqrt{R}\right)^2 \,. 
\end{aligned}
\end{equation}

Solving \eqref{eq:2color:master_theory} for $R$ yields an expression for $R_{\text{opt}}$ depending on the harmonic order corresponding to $E_H$ that is to be enhanced. 
\subsection{Theory - experiment comparison}
\begin{figure}
    \centering
    \includegraphics{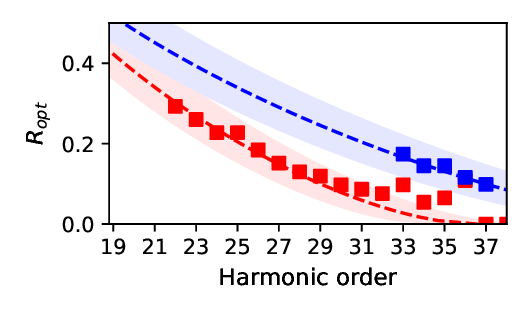}
    \caption{Comparison between \eqref{eq:2color:master_theory} (dashed lines) and experimental results (squares) for $I_{\text{tot}}=\SI{0.9e14}{\watt\per\centi\meter\squared}$ (red) and $I_{\text{tot}}=\SI{1.35e14}{\watt\per\centi\meter\squared}$ (blue). The shaded areas represent the effect of a $\SI{10}{\percent}$ change of $I_{\text{tot}}$ and $R$ in \eqref{eq:2color:master_theory}.}
    \label{fig:comparison}
\end{figure}

In Fig.~\ref{fig:comparison}, \eqref{eq:2color:master_theory} is compared to the experimental results. The theory is shown as a dashed line and the experimentally found $R_{\text{opt}}$  values are indicated as squares, for $I_{\text{tot}}=\SI{0.9e14}{\watt\per\centi\meter\squared}$ (red) and $I_{\text{tot}}=\SI{1.35e14}{\watt\per\centi\meter\squared}$ (blue).
The shaded areas represent the effect of a $\SI{10}{\percent}$ change of $I_{\text{tot}}$ and $R$ in \eqref{eq:2color:master_theory} to highlight the effect of a realistic error margin for optical intensity measurements in an experiment. The red squares hereby correspond to the data shown in Fig.~\ref{fig:results}, for harmonic orders where a global yield maximum could be reached. For $I_{\text{tot}}=\SI{1.35e14}{\watt\per\centi\meter\squared}$, the maximum value for $R$ that could be reached experimentally decreased to $R=0.2$, only a small amount of second harmonic was available due to the design of the interferometer \cite{raabHighlyVersatileTwocolor2024}. As the setup was originally designed for the observation of low-order high harmonics the signal-to-noise ratio after $H37$ decreases rapidly. Within the shown observation range in Fig.~\ref{fig:comparison}, the experimental results show excellent agreement with \eqref{eq:2color:master_theory}.

The relative yield enhancement compared to $R=0$ shown in blue in Fig.~\ref{fig:fit} a) seems to be proportional to $R_{\text{opt}}$ (dashed cyan line). The overall best enhancement according to catastrophe theory should occur at the parameters with the largest degree of singularity, which corresponds to the very flat plateau at $(R,\phi) \approx (0.2,\SI{-1}{\radian})$ (see Fig.~\ref{fig:trajectories1} g)). In terms of overall yield enhancement (sum of all harmonic orders) the experiment agrees as well with the highest yield at $R\approx 0.2$, shown in Fig.~\ref{fig:fit} a).  The relatively increasing individual enhancement below $H23$ can be explained by the very low signal at $R=0$ for these harmonics.

\subsection{Implications for phase-matching}
The discussion so far only considers the single-atom response. Thus, the validity of \eqref{eq:2color:master_theory} is independent of other experimental parameters such as gas target design or pulse duration.
%which were shown to play a large role in propagation effects \cite{constant_optimizing_1999, weissenbilderHow2022}.  
The fairest comparison of XUV yield enhancement that can be reached for a harmonic order would be to compare two perfectly phase-matched experimental conditions. This was already realized in a previous work when investigating the influence of relative  polarization in a two-color setup, achieving an unexpected higher enhancement for perpendicular polarization than for parallel polarization \cite{kim_highly_2005}, contradicting single-atom response calculations \cite{eichmannPolarizationdependentHighorderTwocolor1995}.

In this work, the experimental conditions for ${R=0}$ and ${R\neq0}$ are identical regarding focusing geometries, gas target length, gas pressure, and total intensity to ensure reproducibility. For every $R$ and $\phi$ variation, the phase matching conditions change as the phase-mismatch due to ionization varies. The sensitivity of the XUV conversion efficiency with these changes depends on the gas target length relative to the Rayleigh length of the laser.  

According to Weissenbilder et al., HHG is efficient either for a short gas target with high gas pressure or for a long gas target at lower pressure  \cite{weissenbilderHow2022}. 
In our experiment, the gas target length is estimated to be equal to $\approx \SI{100}{\micro\meter}$, which is short compared to the Rayleigh lengths of $\SI{2.2}{\milli\meter}$ ($\omega$) and $\SI{1.1}{\milli\meter}$ ($2\omega$). This places our experiment in the phase-matching regime which strongly depends on the ionization degree in the medium. 
Assuming that phase-matching is optimized for $R=0$, the enhancement, which is obtained for a given $(R,\phi)$ (see blue axis in Fig.~\ref{fig:fit}) without changing the generation conditions, might become even larger if the interaction parameters are adjusted to optimize phase-matching with the two-color field. Such an investigation will require further experiments and theoretical calculations.

\section{Conclusion}
In this work, we study HHG in argon using the fundamental of an Ytterbium laser and its second harmonic. We measure the harmonic yield in argon as a function of the relative intensity ratio and phase between the two colors to find the optimum generation parameters. The optimum ratio decreases with harmonic order, while the optimum phase between the two fields is independent of the order and equal to $\phi_{\text{opt}} = \SI{-1}{\radian}$.
We explain these findings using an analysis based on the  semi-classical three-step model and propose a simple equation allowing the calculation of the best fraction of second harmonic intensity for the desired harmonic order in the HHG plateau (\eqref{eq:2color:master_theory}). 
This equation depends only on the fundamental wavelength and gas species through the ionization energy and total available intensity. 

\begin{backmatter}
\bmsection{Funding} The authors acknowledge support from the Swedish Research Council (Grant Nos. 2013-8185, 2021-04691, 2022-03519, and 202304603), the European Research Council (advanced grant QPAP, Grant No. 884900), the Crafoord Foundation, and the Knut and Alice Wallenberg Foundation. A.L’H. is partly supported by the Wallenberg Center for Quantum Technology (WACQT), funded by the Knut and Alice Wallenberg Foundation.

%\bmsection{Acknowledgment} We thank some other people as well
\bmsection{Disclosures} The authors declare no conflicts of interest.

\bmsection{Data Availability Statement} 
Data underlying the results presented in this paper are not publicly available at this time but may be obtained from the authors upon reasonable request.

%\bmsection{Supplemental document}
%See Supplement 1 for supporting content.
\end{backmatter}

% Bibliography
\bibliography{paper}

% Full bibliography added automatically for Optics Letters submissions; the following line will simply be ignored if submitting to other journals.
% Note that this extra page will not count against page length
\bibliographyfullrefs{sample}

\end{document}